\documentclass[12pt]{iopart}

\usepackage{iopams}  
\usepackage{graphicx}
\begin{document}

\title{IRGAC2006}{Dark Energy and the False Vacuum}

\author{P Q Hung}

\address{Dept. of Physics, University of Virginia, \\
382 McCormick Road, P. O. Box 400714, Charlottesville, Virginia 22904-4714, 
USA}
\ead{pqh@virginia.edu}
\begin{abstract}
In this talk, I will present highlights of a recent model of dark energy and
dark matter in which the present universe is ``trapped'' in a {\em false vacuum}
described by the potential of an axion-like scalar field (the acceleron) 
which is related to
a new strong interaction gauge sector, $SU(2)_Z$, characterized by a scale
$\Lambda_Z \sim 3 \times 10^{-3}\,eV$. This false vacuum model mimicks
the $\Lambda CDM$ scenario. In addition, there are several additional implications
such as a new mechanism for leptogenesis coming from the decay
of a ``messenger'' scalar field, as well as a new model of ``low-scale'' inflation
whose inflaton is the ``radial'' partner of the acceleron.

\end{abstract}

\submitto{\JPA}
\maketitle

\section{Dark Energy and the False Vacuum}

It is by now customary to present the ``energy budget'' to
illustrate the relative importance of the various components which comprise
the present universe. With $\Omega_X = \rho_X/\rho_c$, one has
$\Omega_{baryons} \sim 4\%$ for baryons (visible and dark),
$\Omega_{DM} \sim 23\%$ for non-baryonic dark matter, and
$\Omega_{DE} \sim 73 \%$ for the mysterious dark energy which
is responsible for the present acceleration of the universe. In terms of energy
density, the latter (dominant) fraction is usually expressed as
$\rho_V \approx (3 \times 10^{-3}\,eV)^4$. Tremendous efforts have
been and will be made to probe the nature of this dark energy. The
latest constraint given in terms of the equation of state of
the dark energy $p = w\, \rho$ gives a value \cite{dark} for $w$,
$w \approx -1$, which is quite consistent
with the $\Lambda CDM$ scenario with $w=-1$. Hopefully, 
the important question
concerning the nature of the dark energy will be resolved by future
projects which could in principle go to high redshifts and determine
whether or not the equation of state is varying with $z$.

If the present universe appears to be one which is dominated by a cosmological
constant $\Lambda$, we are faced with a very uncomfortable question: Why is it
so {\em small}, i.e. why is $\rho_V \sim 10^{-123} M_{pl}^4$? This is the 
``new'' cosmological constant problem
as compared with the ``old'' cosmological constant problem which is one
in which one searches for a reason why it should be exactly equal to zero.
If indeed there is such a {\em reason} then the present value of the cosmological
constant (or something that mimicks it) should be considered to be just
a ``transient'' phenomenon with the universe being stuck in some kind
of false vacuum which will eventually decay into the {\em true vacuum} with a
{\em vanishing} cosmological constant. In this case, the problem boils down to the
search for a {\em dynamical} model in which the false vacuum energy
density is $\sim (3 \times 10^{-3}\,eV)^4$. Furthermore, such a  {\em reason}
would {\em prevent} the existence of any {\em remnant} of vacuum energies
associated with various spontaneous symmetry breakdowns (SSB) (Electroweak,
QCD, and possibly others). For example, it would prevent a
partial cancellation of the electroweak vacuum energy down to the present
value since that would constitute a fundamental cosmological constant in
contradiction with that premise. The true electroweak vacuum would then
have $\Lambda=0$. And similarly for other ({\em completed}) phase transitions.
Anything that mimicks a non-zero cosmological constant would be associated
with a false vacuum. What could then be this sought-after deep reason
for the cosmological constant to vanish in a true vacuum? Needless to say, this is
a fundamental and very difficult question and 
there are many interesting approaches 
for tackling it. It is beyond the scope of this talk to discuss all of them.
One recent interesting proposal \cite{mersini} dealt with
the consequences of the existence of a fundamental cosmological constant. It
was argued in \cite{mersini} that, within the framework of general relativity,
catastrophic gravitational instabilities which are developed during the DeSitter Epoch
(for a fundamental $\Lambda$) would reverse the arrow of time disagreeing
with observations and leading the author to conclude that either one forbids
a fundamental cosmological constant or one modifies general relativity during
the epoch dominated by that constant. We will adopt the former point of view,
namely a vanishing cosmological constant for the true vacuum.

In what follows, I will describe 
a model \cite{hung1}, \cite{hung2} based on the assumption 
that the true vacuum has a vanishing cosmological constant and 
that we are presently trapped in a false vacuum
with an energy density $\rho_V \approx (3 \times 10^{-3}\,eV)^4$. I will argue that
the value $\sim 10^{-3}\,eV$ represents a {\em new dynamical scale} associated with a new gauge group $SU(2)_Z$ \cite{goldberg} which grows strong at that scale. 
In this model, the present
acceleration of the universe is driven by an axion-like particle denoted
by $a_Z$ whose potential is induced by $SU(2)_Z$ instanton effects and which
exhibits two minima: the false vacuum in which $a_Z \neq 0$
and $\rho_V \approx (3 \times 10^{-3}\,eV)^4$, and the true vacuum
in which $a_Z=0$ and $\rho_V =0$.  One of the important features
of this model is that it can be {\em testable} in future collider (such as
the LHC) experiments. This is because the model contains a scalar field-
the so-called messenger field- with a mass less than 1 TeV and which carries
both $SU(2)_Z$ and electroweak quantum numbers. This
and other consequences will be discussed below.

First I will briefly describe the model with its particle content
as well as its results. Next I will describe in a little more detail
what these results mean.

\begin{itemize}

\item The model in \cite{hung1}, \cite{hung2} is based on an {\em unbroken}
vector-like gauge group $SU(2)_Z$. This group contains 
fermions, $\psi^{(Z)}_{i}$ with $i=1,2$,
which transform as a {\em triplets} under $SU(2)_Z$ and as {\em singlets}
under the SM, as well as ``messenger'' scalar
fields, $\tilde{\varphi}_{1,2}^{(Z)}$, which carry both quantum numbers: 
a {\em triplet} under $SU(2)_Z$
and a {\em doublet} under $SU(2)_L$. In addition, there is a complex
singlet (under both sectors) scalar field $\phi_Z = 
(\sigma_Z+v_{Z})\,\exp(ia_Z/v_{Z})$ which
couples only to $\psi^{(Z)}_{i}$ because of a global $U(1)_{A}^{(Z)}$
symmetry. 

\item $\langle \phi_Z \rangle = v_Z$ spontaneously breaks the
$U(1)_{A}^{(Z)}$ symmetry with $a_Z$ becoming a pseudo-Nambu-Goldstone
boson (PNGB) because of the explicit breaking due to $SU(2)_Z$ instanton
effects. Notice that $a_Z$ is very similar to the Peccei-Quinn axion \cite{axion}
in QCD except that we are dealing with another gauge group at another
scale. It is $a_Z$ which plays the role of the {\em acceleron} in our
model \cite{hung2}. And it is also $\sigma_Z$ that plays the role
of the {\em inflaton} in a ``low scale'' inflationary scenario \cite{barcelona}. 

\item The potential $V(a_Z)$ which plays a crucial role in the
dark energy aspect of the model is induced by $SU(2)_Z$ instanton
effects which become more relevant as the gauge coupling grows larger.
In order for the  $SU(2)_Z$ coupling $\alpha_Z = g_Z^2/4\,\pi \sim
1$ at $\Lambda_Z \approx 3 \times 10^{-3}\,eV$, it was found that
a number of constraints had to be satisfied (all of which have
further implications): (1) the initial coupling at high energies has
a value of the order of the SM couplings at comparable energies; (2)
the masses of the $SU(2)_Z$ fermions $\psi^{(Z)}_{i}$ are in the range
of $100-200\,GeV$ and that of the lightest of the messenger field
$\tilde{\varphi}_{1}^{(Z)}$ being in the range $300-1000\,GeV$. One
may ask at this point why $\alpha_Z$ would be of the order of the SM couplings 
at high energies. It turns out that $SU(2)_Z$ can be ``grand unified''
with the SM into the gauge group $E_6$ \cite{hung3} 
which however breaks down quite differently from the usual approach:
$E_6 \rightarrow
SU(2)_Z \otimes SU(6) \rightarrow SU(2)_Z \otimes 
SU(3)_c \otimes SU(3)_L \otimes U(1)
\rightarrow SU(2)_Z \otimes SU(3)_c \otimes SU(2)_L \otimes U(1)_Y
\rightarrow SU(2)_Z \otimes SU(3)_c \otimes U(1)_{em}$.

\item With the value of the $SU(2)_Z$ gauge coupling at
a temperature of $O(200\,GeV)$ (the
favored mass range for the fermions $\psi^{(Z)}_{i}$)
being of the order of the electroweak coupling, its annihilation cross
section was found to be typically of the order of a weak cross section
and thus providing ideal (WIMP) cold dark matter candidates in the
form of $\psi^{(Z)}_{i}$ \cite{hung2}.

\item The lighter of
the two messenger fields, $\tilde{\varphi}_{1}^{(Z)}$, which 
carries both $SU(2)_Z$ and electroweak quantum numbers, can couple 
only to $\psi^{(Z)}_{i}$ and a SM lepton. Its decay in the early
universe can generate a SM lepton number asymmetry which transmogifies
into a baryon number asymmetry through electroweak sphaleron processes
\cite{hung4}.

\end{itemize}

Basically, the $SU(2)_Z$ instanton-induced potential $V(a_Z)$ has two degenerate
vacuua due to the remaining $Z(2)$ symmetry (2 ``flavors'' of $\psi^{(Z)}_{i}$),
and is expressed as $V(a_Z,T) = \Lambda_Z^4[1-\kappa(T)\,\cos\frac{a_Z}{v_Z}]$,
where $\kappa (T) =1$ at $T=0$. This is lifted by a soft-breaking term
$\kappa (T)\Lambda_Z^4\,\frac{a_Z}{2\pi\,v_Z}$ which is linked to $SU(2)_Z$ fermion
condensates \cite{hung5}. This is shown in the following figure 
for $V(a_Z,T)/\Lambda_Z^4$ as a function of $a_Z/v_Z$ and for
$T \ll \Lambda_Z$:

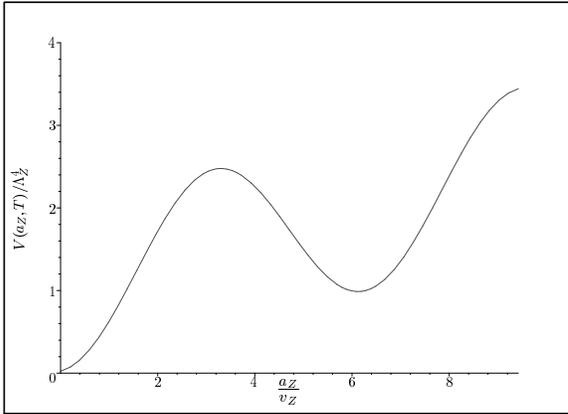
\includegraphics[angle=-90,width=3in]{V_Z2p.epsi}

From the above figure, one notices that the metastable (false) vacuum is
at $a_Z  = 2 \pi v_Z$ while the true vacuum is at $a_Z =0$.
For $T \gg \Lambda_Z$, $V(a_Z,T)$ is relatively flat because
$SU(2)_Z$ isntanton effects are negligible there. One also expects
$a_Z$ to hover around $O(v_Z)$. It is assumed that,
as $T < \Lambda_Z$, the universe 
got trapped in the false vacuum with an energy density
$\rho_V = \Lambda_Z^4 \approx (3 \times 10^{-3}\,eV)^4$. 

It is interesting to estimate the various ages of the universe in this scenario:
(1) Age of the universe when the $SU(2)_Z$ coupling grows strong 
($\alpha_Z =1$) at $T_Z \sim
3 \times 10^{-3}\,eV \sim 35 ^{0}K$ corresponding to the background
radiation temperature $T \approx 70 ^{0}K$: $z \approx 25$,
$t_z \approx 125 \pm 14 Myr$;
(2) Age of the universe when the deceleration ``stopped''
and the acceleration ``started'' ($\ddot{a}=0$): $z_a \sim 0.67$,
$t_a \approx 7.2 \pm 0.8\,Gyr$; (3) Age of the universe when
$\rho_M \sim \rho_{V}$: $z_{eq} \approx 0.33$, $t_{eq} = 9.5 \pm 1.1 \,Gyr$.

Notice that the equation of state is
$w = \frac{p}{\rho}=\frac{\frac{1}{2}\dot{a_Z}^2 - V(a_Z)}
{\frac{1}{2}\dot{a_Z}^2 + V(a_Z)} < 0$ for $\frac{1}{2}\dot{a_Z}^2 \ll V(a_Z)$.
With the present universe being trapped in a false vacuum,
$\frac{1}{2}\dot{a_Z}^2 \sim 0$ leading to $w \approx -1$.
Our model effectively mimics the $\Lambda CDM$ scenario.

How long will it take for the false vacuum $a_Z  = 2 \pi v_Z$ to make a transition
to the true vacuum $a_Z =0$? A rough estimate using the thin wall aproximation
gives a bound on the Euclidean action 
$S_E \geq 5 \times 10^{5}\,(\frac{v_Z}{\Lambda_Z})^4 \geq 10^{89}$ for $v_Z
\sim 10^{9}\, GeV$ (as deduced from the low-scale inflation model \cite{barcelona}).
With the bubble nucleation rate $\Gamma = A \exp\{-S_E\}$ ($A \sim O(1)$)
and the transition time $\tau = \frac{3\,H}{4\,\pi \Gamma} \geq 
(10^{-106}\,s) \exp(10^{89})$, one can see that indeed it would take a {\em very long}
time for this to occur. As anticipated by many people, the universe will enter an
inflationary stage and, in this scenario, the ``late'' inflation will last an
``astronomical'' time. Although it is entirely academic, it is interesting to note
that the ``reheating'', after this late inflation stops, occurs through the decay
of $a_Z$ into two $SU(2)_Z$ ``gluons'' which, in turns, produce the messenger
field  $\tilde{\varphi}_{1}^{(Z)}$ 
and eventually SM leptons followed by SM quarks. (A somewhat analogous reheating
mechanism for the low-scale (early) inflation \cite{barcelona} was also proposed.)

\section{Implications of the dark energy model}

I) The first implication of this scenario is the existence of possible
candidates for WIMP-like Cold Dark Matter in the form of $\psi^{(Z)}_{i}$.
Notice that $\psi^{(Z)}_{1,2} = (3,1)$ under $SU(2)_Z \otimes SM$ and
have a mass $\sim$ O($100-200\,GeV$).
The condition for $\psi^{(Z)}$ to be CDM
candidates is
$\Omega_{\psi^{(Z)}} = \frac{m_{\psi^{(Z)}}\,n_{\psi^{(Z)}}}
{\rho_{c}} \approx \bigl(\frac{3 \times 10^{-27}\,cm^{3}\,sec^{-1}}
{\langle \sigma_{A,\psi^{(Z)}}\,v\rangle} \bigr)$,
with the annihilation cross section $\langle \sigma_{A,\psi^{(Z)}}\rangle$ being 
typically of the order of a weak cross section, i.e. $\langle 
\sigma_{A,\psi^{(Z)}}\rangle \sim 10^{-36}\, cm^2 \sim \frac{3 \times
10^{-9}}{GeV^2}$ in order for $\Omega_{\psi^{(Z)}} \sim O(1)$. This 
is the so-called WIMP.
It was noticed in \cite{hung1} and \cite{hung2} that $\psi^{(Z)}$ with 
a mass $\sim$ O($100-200\,GeV$) would do just that since
one expects $\langle 
\sigma_{A,\psi^{(Z)}}\rangle \sim \frac{\alpha_{Z}(T)^2}{m_{\psi^{(Z)}}^2}$
and $\alpha_{Z}(T)^2 \sim 6 \times 10^{-4}$ over a large range of energy
down to $\sim 100\,GeV$.

How do we detect those CDM candidates? The most obvious way would be an indirect
method: $\tilde{\varphi}_{1}^{(Z)} \rightarrow \bar{\psi}^{(Z)}_{1,2}+ l$, where
$l$ stands for a SM lepton. A pair of $\tilde{\varphi}_{1}^{(Z)}$ could be produced
at the LHC through electroweak gauge boson fusion processes. The decays would have
unusual geometries (e.g. the SM leptons need not be back-to-back) and $\psi^{(Z)}$
would ``appear'' as missing energies.

II) The second implication concerns a new mechanism for Leptogenesis
via the decay of a ``messenger'' scalar field 
$\tilde{\varphi}_{1}^{(Z)}
= (3,1,2,Y/2=-1/2)$ under $SU(2)_Z \otimes SU(3)_c \otimes SU(2)_L \otimes U(1)_Y$.
As discussed in \cite{hung4}, the asymmetry between
$\tilde{\varphi}_{1}^{(Z)} \rightarrow \bar{\psi}^{(Z)}_{1,2}+ l$
and $\tilde{\varphi}_{1}^{(Z),*} \rightarrow \psi^{(Z)}_{1,2}+ \bar{l}$ could provide
a {\em net SM lepton number}. This becomes a
net baryon number through EW sphaleron processes. It is by now a familiar phenomenon
that the asymmetry comes from the interference between tree-level and one-loop 
contributions to the decays. Also, for the asymmetry $\neq 0$, we need {\em two} 
messenger fields: $\tilde{\varphi}_{1,2}^{(Z)}$, with 
$m_{\tilde{\varphi}_{2}^{(Z)}} \gg
m_{\tilde{\varphi}_{1}^{(Z)}}$. The asymmetry which is defined as
$\epsilon^{\tilde{\varphi}_{1}} = 
(\Gamma_{\tilde{\varphi}_{1}\,l}-
\Gamma_{\tilde{\varphi}_{1}^{*}\,\bar{l}})/(\Gamma_{\tilde{\varphi}_{1}\,l}+
\Gamma_{\tilde{\varphi}_{1}^{*}\,\bar{l}})$ is roughly $-10^{-7}$. 
This estimate comes from the SM lepton number asymmetry ($n_{LSM}$) 
per unit entropy ($s$):
$n_{LSM}/s \sim 2 \times 10^{-3}\,
\epsilon^{\tilde{\varphi}_{1}}_{l}$, which, in turns, is related to the
baryon number per unit entropy
$n_{B}/s \sim -0.35\,n_{LSM}/s\,\sim -10^{-3}\,
\epsilon^{\tilde{\varphi}_{1}}_{l} \sim 10^{-10}$, where the coefficient
$-0.35$ is for the SM with three families and one Higgs doublet. 
In \cite{hung3}, it is shown that this
puts an {\em upper} bound on the mass of the messenger field: 
$m_{\tilde{\varphi}_{1}} \leq 1\,TeV$. This makes a search for this ``progenitor
of SM lepton number'', $\tilde{\varphi}_{1}$, fairly feasible at the LHC
if its mass is low enough.

III) The third implication comes from the interesting possibility that 
$\sigma_Z$ ($\phi_Z = (\sigma_Z+v_{Z})\,\exp(ia_Z/v_{Z})$) can play the
role of the inflaton in a ``low-scale'' inflationary scenario \cite{barcelona}.
It was proposed that a Coleman-Weinberg potential for $\sigma_Z$ is
consistent with recent WMAP3 data on the spectral index $n_s$ for
$v_Z \sim  10^9\,GeV$. The inflaton mass $m_{\sigma_Z}\simeq 450 \,GeV$ is low
enough so that it might be indirectly ``observed'' at colliders such as the LHC
through its coupling with $\psi^{(Z)}_{1,2}$ which, in turns, couple to
$\tilde{\varphi}_{1}^{(Z)}$.

IV) The fourth implication is the possibility of unifying $SU(2)_Z$ with the SM
into $E_6$ as mentioned above \cite{hung3}. This unification requires the
existence of heavy mirror fermions which could be searched for at future colliders.
An estimate for the proton lifetime gives, however, a mean value about an order
of magnitude larger than the present lower bound ($\sim 2 \times 10^{32}\,
yrs$) which makes it inaccessible experimentally for quite some time.

\ack This work is supported in parts by the US Department of Energy under grant No.
DE-A505-89ER40518. I would like to thank Joan Sol\`{a} and the organizing committee
for a nice and comprehensive meeting.

\end{document}